\documentclass[aps,prl,twocolumn]{revtex4}
\usepackage{graphicx,amssymb,amsfonts,amsmath,dcolumn,bm}

\begin{document}

\title{Nature of the Bogoliubov ground state of a weakly interacting Bose gas}

\author{A.M. Ettouhami} 


\email{ettouhami@gmail.com}

\affiliation{Department of Physics, University of Toronto, 60 St. George St., 
Toronto M5S 1A7, Ontario, Canada}

\altaffiliation[Present address: ]{20 Antrim Crescent, Apt. 1412, Scarborough, Ontario M1P 4N3, Canada}

\date{\today}

\begin{abstract}

As is well-known, in Bogoliubov's theory of an interacting Bose gas
the ground state of the Hamiltonian $\hat{H}=\sum_{\bf k\neq 0}\hat{H}_{\bf k}$ 
is found by diagonalizing each of the Hamiltonians $\hat{H}_{\bf k}$ corresponding
to a given momentum mode ${\bf k}$ independently of the Hamiltonians $\hat{H}_{\bf k'(\neq k)}$ 
of the remaining modes. We argue that this way of diagonalizing $\hat{H}$ may not be adequate,
since the Hilbert spaces where the single-mode Hamiltonians $\hat{H}_{\bf k}$
are diagonalized are not disjoint, but have the ${\bf k}=0$ in common.
A number-conserving generalization of Bogoliubov's method is presented where the total Hamiltonian 
$\hat{H}$ is diagonalized directly. When this is done,
the spectrum of excitations changes from a gapless one, as predicted by
Bogoliubov's method, to one which has a finite gap in the $k\to 0$ limit.

\end{abstract}

\pacs{03.75.Hh,03.70.+k,03.75.-b}


\keywords{Bose-Einstein condensation; Bogoliubov theory; Interacting Bose gas}

\maketitle

{\em Introduction.} Since its inception in 1947, Bogoliubov's approach to interacting Bose systems \cite{Bogoliubov1947}
has been one of the most influential theories in condensed matter physics. 
\cite{Lee1957,LeeHuangYang1957,Bruckner1957,Beliaev1958,Hugenholtz1959,Sawada1959,Gavoret1964,Hohenberg1965,Popov1965,Singh1967,Cheung1971,Szepfalusy1974}
Yet, for all its notoriety and popularity within the physics community,
a key aspect of this theory, having to do with the decoupled way in which
the Hamiltonian is diagonalized, is still not fully understood. Indeed, and 
as is well-known, in the standard formulation of Bogoliubov's theory,
the Hamiltonian $\hat{H}$ is written as a decoupled sum of contributions from different momenta of the form 
$\hat{H} = \sum_{\bf k\neq 0}\hat{H}_{\bf k}$, where each Hamiltonian $\hat{H}_{\bf k}$
describes the interaction of bosons in the condensed $\bf k = 0$ state 
with bosons in the momentum modes $\pm\bf k$, then each of the single-mode Hamiltonians $\hat{H}_{\bf k}$
is diagonalized {\em separately} and the ground state (GS) wavefunction of $\hat{H}$
is written as the product of the GS wavefunctions of the $\hat{H}_{\bf k}$'s.
In this letter, we shall argue that, 
while this way of diagonalizing the total Hamiltonian
$\hat{H}$ may seem to be valid from the perspective of the standard,
number non-conserving Bogoliubov's method, where the $\bf k=0$ state
is removed from the Hilbert space and hence the individual Hilbert spaces where the Hamiltonians
$\{\hat{H}_{\bf k}\}$ are diagonalized are disjoint with one another, from a number-conserving perspective 
this diagonalization method may not be adequate since the true Hilbert spaces
where the Hamiltonians $\{\hat{H}_{\bf k}\}$ should be diagonalized all have the  
${\bf k}=0$ state in common. We then shall discuss a variational, number-conserving 
generalization of Bogoliubov's theory in which the ${\bf k}=0$ state is restored
into the Hilbert space of the interacting gas, and where, instead of diagonalizing the Hamiltonians
$\hat{H}_{\bf k}$ separately, we diagonalize the total Hamiltonian $\hat{H}$ as a whole. 
When this is done, the spectrum of excitations of the system changes from a gapless one, 
as predicted by the standard, number non-conserving
Bogoliubov method, to one which exhibits a finite gap in the $k\to 0$ limit.

{\em Variational formulation of Bogoliubov's theory.}
We shall start by discussing a variational formulation \cite{LeeHuangYang1957,LeggettRMP}
of Bogoliubov's theory which, historically, has constituted the basis of the justification of the number non-conserving formulation
of this method. As is well-known, in Bogoliubov's approach 
one only retains in the total Hamiltonian $\hat{H}$ of the system
{\it (i)} kinetic energy terms of the form $\sum_{\bf k\neq 0}\varepsilon_{\bf k}a_{\bf k}^\dagger a_{\bf k}$,
{\it (ii)} Hartree terms $\sum_{\bf k,k'}v({\bf 0})a_{\bf k}^\dagger a_{\bf k'}^\dagger a_{\bf k}a_{\bf k'}/2V$, 
{\it (iii)} Fock terms $\sum_{\bf k\neq 0} v({\bf k}) a_{\bf k}^\dagger a_{\bf k} a_{\bf 0}^\dagger a_{\bf 0}/V$
describing the exchange interaction between condensed bosons and depleted ones, and 
{\it (iv)} pairing terms of the form $\sum_{\bf k\neq 0}v({\bf k})(a_0 a_0 a_{\bf k}^\dagger a_{-\bf k}^\dagger + 
a_0^\dagger a_0^\dagger a_{\bf k} a_{-\bf k})/2V$. (In the above expressions, 
$\varepsilon_{\bf k}=\hbar^2k^2/2m$ is the kinetic energy of a boson of mass $m$ and wavevector ${\bf k}$,
$v({\bf k})$ is the Fourier transform of the interaction potential between bosons, and
$V$ is the volume of the system. On the other hand, $a_{\bf k}^\dagger$ and $a_{\bf k}$
are creation and annihilation operators, respectively). Considering that we will be
focusing on systems having a fixed number of particles $N$, it is convenient to 
take the origin of energies at the Gross-Pitaevskii value $v(0)N(N-1)/2V$. Then
it can be shown \cite{LeggettRMP}
that the Hamiltonian can be written as a sum of independent contributions from different values
of ${\bf k}$ of the form $\hat{H} \simeq \sum_{\bf k \neq 0} \hat{H}_{\bf k}$,
where (throughout this paper, for all explicit calculations
we shall be using the interaction potential $v({\bf r})=g\delta({\bf r})$, for which $v({\bf k})=g$):
\begin{align}
\hat{H}_{\bf k} = \varepsilon_{\bf k}a_{\bf k}^\dagger a_{\bf k} 
& + \frac{v({\bf k})}{2V}\,\big(2a_0^\dagger a_0 a_{\bf k}^\dagger a_{\bf k} 
\nonumber\\
& + a_{\bf k}^\dagger a_{-\bf k}^\dagger a_0 a_0
+ a_0^\dagger a_0^\dagger a_{\bf k} a_{-\bf k}\big).
\label{Eq:defHk}
\end{align}
We now proceed to diagonalize the Hamiltonian $\hat{H}_{\bf k}$ by considering
a hypothetical system where bosons are only allowed to be 
in one of the three single particle states with momentum ${\bf k}$, ${\bf 0}$ or $-\bf k$.
In order to formulate a variational approach for the Hamiltonian $\hat{H}_{\bf k}$ describing such
a system, it is sufficient to restrict our attention to the Hilbert space ${\cal H}_{\bf k}$ 
spanned by kets $|n\rangle$ of the form:
\begin{equation}
|n\rangle \equiv |N-2n,n,n\rangle = \frac{(a_0^\dagger)^{N-2n}}{\sqrt{(N-2n)!}}\frac{(a_{\bf k}^\dagger)^n}{\sqrt{n!}}
\frac{(a_{-\bf k}^\dagger)^n}{\sqrt{n!}}|0\rangle,
\label{Eq:Hilbert}
\end{equation}
having $n$ bosons with momentum ${\bf k}$ and momentum $-{\bf k}$,
and $N-2n$ bosons in the ${\bf k}={\bf 0}$ state.
The general expression of the GS wavefunction $|\psi_{\bf k}\rangle$ 
of the Hamiltonian $\hat{H}_{\bf k}$ in this Hilbert space is given by
$|\psi_{\bf k}\rangle = \sum_{n=0}^{N/2} C_n|n\rangle$,
and it can easily be verified that the expectation value of $\hat{H}_{\bf k}$ 
in the state $|\psi_{\bf k}\rangle$ can be written in the form:
\begin{align}
&\langle\psi_{\bf k}|\hat{H}_{\bf k}|\psi_{\bf k}\rangle = \sum_{n=0}^{N/2}\Big\{
|C_n|^2\Big[n\varepsilon_{\bf k} + \frac{v(k)}{V}n(N-2n)\Big]
\nonumber\\
&+\frac{v(k)}{2V}nC_n^*C_{n-1}\sqrt{(N-2n+2)(N-2n+1)}
\nonumber\\
&+\frac{v(k)}{2V}(n+1)C_n^*C_{n+1}\sqrt{(N-2n-1)(N-2n)}
\Big\},
\label{Eq:PsihPsi1}
\end{align}
where it is understood that $C_{-1}=C_{1+(N/2)}=0$.
Assuming, for simplicity, that the coefficients $C_n$ are real, 
it follows that, for $v(k)>0$, the expectation value 
$\langle\psi_{\bf k}|\hat{H}_{\bf k}|\psi_{\bf k}\rangle$
will be lowered if the coefficients $C_n$ have alternating positive and 
negative signs. In this case, the terms on the second and third line will be negative, 
making the expectation value lower than what one would obtain if products of the form
$C_nC_{n-1}$ and $C_nC_{n+1}$ are positive. 
Bogoliubov's theory corresponds to a variational ansatz in which the coefficients $C_n$
are assumed to be of the form $C_n = (-c_{\bf k})^n$,
where the constant $c_{\bf k}>0$ is to be determined variationally. 
The coefficients $C_n$ are expected to decrease with $n$, 
which encodes the fact that the probability amplitude of states
$|n\rangle$ with a large number $n\gg 1$ of bosons having a wavevector $\pm{\bf k}$ will be small. This implies
that the constant $c_{\bf k}$ must be less than unity.

\begin{figure}[tb]
\includegraphics[width=8.09cm, height=5.5cm]{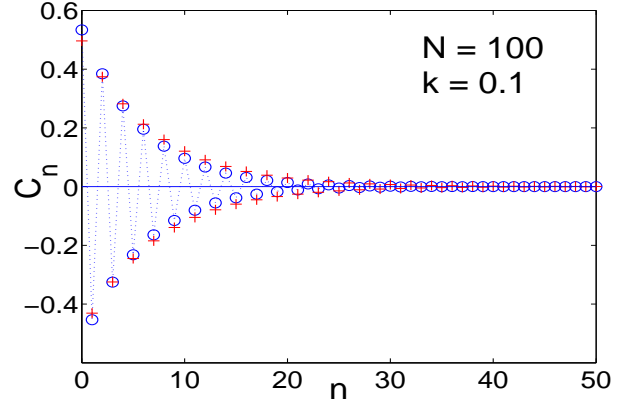}
\caption[]{Coefficients $\widetilde{C}_n$ of the normalized wavefunction 
$|\widetilde\psi_{\bf k}\rangle$ for $N=100$ bosons and $\tilde{k}=0.1$. 
The crosses are the results of Bogoliubov's theory,
and the circles are the results of the exact numerical diagonalization
of the Hamiltonian $\hat{H}_{\bf k}$.
}\label{Fig:CoeffPsiOneMode}
\end{figure}

Inserting the variational ansatz $C_n = (-c_{\bf k})^n$ into Eq. (\ref{Eq:PsihPsi1}),
and making use of the approximation
$\sqrt{N(N+1)}\simeq N + \frac{1}{2}$
which is valid for $N\gg 1$, we can write:
\begin{align}
&\langle\psi_{\bf k}|\hat{H}_{\bf k}|\psi_{\bf k}\rangle \simeq \sum_{n=0}^{N/2}\Big\{
(c_{\bf k})^{2n}\big[n\varepsilon_{\bf k} + \frac{v(k)}{V}n(N-2n)\big]
\nonumber\\
&+\frac{v(k)}{2V}n(c_{\bf k})^{2n-1}\Big(N-2n+\frac{3}{2}\Big)
\nonumber\\
&+\frac{v(k)}{2V}(n+1)(c_{\bf k})^{2n+1}\Big(N-2n-\frac{1}{2}\Big)
\Big\}.
\label{Eq:PsihPsi2}
\end{align}
The summations in the above equation can be calculated analytically 
by taking successive derivatives with respect to the variable $x$ of the 
result $\sum_{n=0}^{N/2} x^n \simeq 1/(1-x)$, valid
for $|x|<1$ and $N\to \infty$, hence obtaining:
\begin{align}
\sum_{n=1}^{N/2} nx^n \simeq \frac{x}{(1-x)^2}, 
\quad
\sum_{n=1}^{N/2} n^2x^n \simeq \frac{x+x^2}{(1-x)^3}. 
\label{Eq:n2xn_asymptotic}
\end{align}
\label{Eq:asymptoticseries}
Using these last two results in Eq. (\ref{Eq:PsihPsi2}), we obtain:
\begin{equation}
\langle\psi_{\bf k}|\hat{H}_{\bf k}|\psi_{\bf k}\rangle =\frac{c_{\bf k}}{(1-c_{\bf k}^2)^2}\Big[
c_{\bf k}\varepsilon_{\bf k} 
+ v(k)n_B\big(c_{\bf k}-1\big)
\Big],
\label{Eq:PsihPsi3}
\end{equation}
where we denote by $n_B=N/V$ the density of bosons in the system.
On the other hand, from the definition of $|\psi_{\bf k}\rangle$, we easily see that the norm of the wavefunction 
$\langle\psi_{\bf k}|\psi_{\bf k}\rangle$ is given by
$\langle\psi_{\bf k}|\psi_{\bf k}\rangle = \sum_{n=0}^{N/2}|C_n|^2 \simeq 1/(1-c_{\bf k}^2)$.
Now, if we divide Eq. (\ref{Eq:PsihPsi3}) by this last expression of $\langle\psi_{\bf k}|\psi_{\bf k}\rangle$, 
we can write for the normalized expectation value $\langle \hat{H}_{\bf k}\rangle_{\bf k} 
= \langle \psi_{\bf k}|\hat{H}_{\bf k}|\psi_{\bf k}\rangle/\langle\psi_{\bf k}|\psi_{\bf k}\rangle$
the following result:
\begin{align}
\langle\hat{H}_{\bf k}\rangle_{\bf k} \simeq 
\frac{c_{\bf k}^2}{1-c_{\bf k}^2}\big[\varepsilon_{\bf k}+v(k)n_B\big] - v(k)n_B\frac{c_{\bf k}}{1-c_{\bf k}^2}.
\label{Eq:expectation_h_min}
\end{align}
Minimization of the above expectation value with respect to $c_{\bf k}$
leads to the quadratic equation
$c_{\bf k}^2 - 2\left(\frac{\cal{E}_{\bf k}}{v(k)n_B}\right) c_{\bf k} + 1 = 0$,
where we defined ${\cal{E}_{\bf k}}=\varepsilon_{\bf k} + v(k)n_B$.
The above quadratic equation has two roots, of which only one satisfies
the constraint $0<c_{\bf k}<1$ for arbitrary
values of ${\cal E}_{\bf k}$. This root is given by:\cite{LeggettRMP}
\begin{equation}
c_{\bf k} = \frac{1}{v(k)n_B}\left[
{\cal E}_{\bf k} - \sqrt{{\cal E}_{\bf k}^2-v(k)^2n_B^2}\right].
\label{Eq:resultck}
\end{equation}
This result for the constant $c_{\bf k}$ fully determines the coeffcients $C_n=(-c_{\bf k})^n$
of the variational ground state $|\psi_{\bf k}\rangle$ of the Hamiltonian $\hat{H}_{\bf k}$.
The coefficients $\widetilde{C}_n = C_n\sqrt{1-c_{\bf k}^2}$ of the normalized wavefunction 
$|\widetilde\psi_{\bf k}\rangle = |\psi_{\bf k}\rangle/\sqrt{\langle\psi_{\bf k}|\psi_{\bf k}\rangle}$
are plotted (as the crosses) in Fig. \ref{Fig:CoeffPsiOneMode} for $\tilde{k}=0.1$ in dimensionless units
such that $\tilde{k}\equiv \hbar k/\sqrt{2mv(0)n_B}$. The circles in this last figure are
the coefficients obtained by direct numerical diagonalization of the Hamiltonian $\hat{H}_{\bf k}$.
It is seen that there is a pretty good agreement between the results of our variational method and
the exact numerical diagonalization for the particular value of $k$ chosen.

The expectation value of $\hat{H}_{\bf k}$ in the ground state $|\psi_{\bf k}\rangle$ can readily be found
if we use the result (\ref{Eq:resultck}) for $c_{\bf k}$ in Eq. (\ref{Eq:expectation_h_min}), 
upon which we obtain:
\begin{align}
\langle\hat{H}_{\bf k}\rangle_{\bf k} = \frac{1}{2}\Big[
\sqrt{\varepsilon_{\bf k}\big[\varepsilon_{\bf k} + 2n_Bv(k)\big]}
-\varepsilon_{\bf k} - n_Bv(k)\Big].
\label{Eq:avgHk}
\end{align}
The result (\ref{Eq:avgHk}) is exactly what one obtains in the standard, number non-conserving
Bogoliubov approach for the expectation value of a given contribution $\hat{H}_{\bf k}$ to
the Bogoliubov ground state energy. This agrees with the well-known fact that Bogoliubov's theory 
is a theory in which the Hamiltonians $\hat{H}_{\bf k}$ are diagonalized independently
from one another in essentially disjoint Hilbert spaces.
Indeed, the quantity in Eq. (\ref{Eq:avgHk}) is nothing but
the expectation value $\langle\widetilde\psi_{\bf k}(N)|\hat{H}_{\bf k}|\widetilde\psi_{\bf k}(N)\rangle$,
where $|\widetilde\psi_{\bf k}(N)\rangle$ is the normalized ground state
{\em of the single momentum mode Hamiltonian} $\hat{H}_{\bf k}$.
The reason such a result is obtained is because in the standard formulation of Bogoliubov's theory,
$a_0$ and $a_0^\dagger$ are replaced by the c-number $\sqrt{N}$. This implies that the commutators
$[\hat{H}_{\bf k},\hat{H}_{\bf k'}]$ vanish identically for ${\bf k}\neq {\bf k'}$, which
allows the ground state wavefunction $|\Psi_B(N)\rangle$
of the total Hamiltonian $\hat{H}$ to be written as a product of the ground state wavefunctions 
for each of the Hamiltonians $\hat{H}_{\bf k}$ \cite{LeeHuangYang1957}
($M$ here denotes the total number of momentum
modes kept in the calculation, which can eventually be taken to infinity):
\begin{align}
&|\Psi_B(N)\rangle \equiv \sum_{n_1=0}^\infty \ldots\sum_{n_M=0}^\infty
\widetilde{C}_{n_1}\widetilde{C}_{n_2}\cdots\widetilde{C}_{n_M}
|n_1;\ldots;n_M\rangle,
\nonumber\\
&\mbox{with}\quad 
|n_1;\ldots;n_M\rangle  = \prod_{i=1}^M \frac{\big(a_{{\bf k}_i}^\dagger)^{n_i}}{\sqrt{n_i!}}
\frac{\big(a_{{-\bf k}_i}^\dagger)^{n_i}}{\sqrt{n_i!}}|0\rangle.
\label{Eq:fullPsi(N)0}
\end{align}
Here, we would like to emphasize that the above expression
of the wavefunction $|\Psi_B(N)\rangle$ is only consistent with the variational constants given 
in Eq. (\ref{Eq:resultck}) when the ${\bf k}=0$ state is removed from the Hilbert space,
with $a_0$ and $a_0^\dagger$ being replaced with $\sqrt{N}$.
This means that the ground state wavefunction of Eq. (\ref{Eq:fullPsi(N)0}) above, 
despite the appearance of the contrary, corresponds to a number {\em non}-conserving approach.
A major question that arises is to know how the above result will
change if we restore the ${\bf k}=0$ state to the Hilbert space, 
and if instead of diagonalizing each of the Hamiltonians $\hat{H}_{\bf k}$ separately,  
we diagonalize $\hat{H}$ directly. This will be done next.

\begin{figure}[tb]
\includegraphics[width=8.09cm, height=5.5cm]{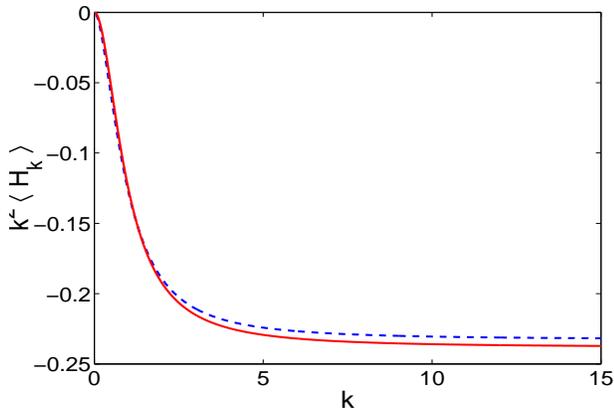}
\caption[]{
Plot of the product 
$\tilde{k}^2\times\langle\Psi|\hat{H}_{\bf k}|\Psi\rangle$ vs. $\tilde{k}$ from Eq. (\ref{Eq:expectation_Hk}),
using the coefficients $c_{\bf k}$
(i) from the single-mode theory, Eq. (\ref{Eq:resultck}), (dashed line) 
and (ii) from Eq. (\ref{Eq:solck}) with $\tilde\sigma=0.39$ and $C_d=0.9762$ (solid line).
}\label{Fig:ktimesGSEnergy}
\end{figure}

{\em Variational approach for the full Hamiltonian $\hat{H}$.} 
We now want to generalize the variational treatment of the single-mode Hamiltonian $\hat{H}_{\bf k}$ to treat
the {\em full} Hamiltonian $\hat{H}=\sum_{\bf k\neq 0}\hat{H}_{\bf k}$ of the interacting Bose system.
To this end, we shall use for $|\Psi(N)\rangle$ the expression:
\begin{align}
|\Psi(N)\rangle & = \sum_{n_1=0}^{\infty} \ldots\sum_{n_M=0}^{\infty}
\widetilde{C}_{n_1}\widetilde{C}_{n_2}\ldots \widetilde{C}_{n_M}
\nonumber\\
&\times |N-2\sum_{i=1}^M n_i; n_1;\ldots;n_M\rangle,
\label{Eq:fullPsi(N)}
\end{align}
where the normalized basis wavefunctions are given by 
(compare with Eq. (\ref{Eq:fullPsi(N)0}) of the single-mode theory):
\begin{align}
|N-2\sum_{i=1}^M n_i;& n_1;\ldots;n_M\rangle  = 
\frac{\big(a_0^\dagger\big)^{N-2\sum_{i=1}^M n_i}}{\sqrt{[N-2\sum_{i=1}^M n_i]!}}
\nonumber\\
&\times\prod_{i=1}^M \frac{\big(a_{{\bf k}_i}^\dagger)^{n_i}}{\sqrt{n_i!}}
\frac{\big(a_{{-\bf k}_i}^\dagger)^{n_i}}{\sqrt{n_i!}}|0\rangle.
\label{Eq:varBasis}
\end{align}
Note that the GS wavefunction in Eq. (\ref{Eq:fullPsi(N)}) is {\em not} a simple product of 
GS wavefunctions for the single-mode Hamiltonians $\hat{H}_{\bf k}$, and that, even
though the expression of these single-mode Hamiltonians $\hat{H}_{\bf k}$ are decoupled and commute with one another,
the presence of all the $n_i$'s in the number of condensed bosons $\big[N-2\sum_{i=1}^Mn_i\big]$
acts like an implicit and rather nontrivial coupling between all these Hamiltonians.
One can now show \cite{Ettouhami} that the expectation value $\langle \hat{H}_{{\bf k}_j}\rangle$
of the Hamiltonian $\hat{H}_{{\bf k}_j}$ in the state $|\Psi(N)\rangle$
is no longer given by Eq. (\ref{Eq:expectation_h_min}), but by the following expression:
\begin{equation}
\langle\hat{H}_{{\bf k}_j}\rangle  \simeq 
\frac{c_{{\bf k}_j}}{1-c_{{\bf k}_j}^2}\Big\{
c_{{\bf k}_j}\Big[\varepsilon_{{\bf k}_j}+\bar{v}(k_j)n_B\Big] 
- \bar{v}(k_j)n_B
\Big\},
\label{Eq:expectation_Hk}
\end{equation}
where $\bar{v}(k_j)$ is given by:
\begin{align}
\bar{v}(k_j) \simeq v(k_j)\Big( 1 - \frac{1}{N}\sum_{\bf k\neq 0}
\frac{c_{{\bf k}}^2}{1-c_{{\bf k}}^2}\Big).
\label{Eq:defbarv}
\end{align}
If it were not for the term between parenthesis in this last equation, the result in Eq. (\ref{Eq:expectation_Hk})
would be perfectly identical to the expectation value obtained within the single-mode approach, 
Eq. (\ref{Eq:expectation_h_min}). 
It can be shown \cite{Ettouhami} that minimization of the trial ground state energy
given in Eq. (\ref{Eq:expectation_Hk}) over the constants $c_{\bf k}$ leads to a solution of the form:
\begin{align}
c_{\bf k} = 1 + C_d^{-1}(\tilde{k}^2 + \tilde\sigma)
- \sqrt{\big[ 1 + C_d^{-1}(\tilde{k}^2 + \tilde\sigma) \big]^2 - 1}.
\label{Eq:solck}
\end{align}
where the constant $\tilde\sigma$ is obtained by solving a non-linear self-consistency equation
obtained from the minimization procedure, \cite{Ettouhami}
and $C_d=1-N_d/N$, with $N_d=\sum_{\bf k\neq 0}c_{\bf k}^2/(1-c_{\bf k}^2)$ the total number of depleted bosons.
To fix ideas, we shall henceforth consider an interacting Bose gas in the dilute limit, and fix
the parameter $n_Ba^3$ to be $n_Ba^3 = 10^{-3}$ ($a$ here being the scattering length, which is related
to the interaction strength $g$ through the relation $g = {4\pi a\hbar^2}/{m} $). 
For this particular value of $n_Ba^3$, we find $\tilde{\sigma}=0.39$ and $C_d =0.9762$. 
In order to show that these values of $\tilde{\sigma}$ and $C_d$
do indeed correspond to a lower energy than what one would obtain by using 
the coefficients $c_{\bf k}$ of the single-mode theory from Eq. (\ref{Eq:resultck}) 
in Eq. (\ref{Eq:expectation_Hk}), in Fig. \ref{Fig:ktimesGSEnergy} we plot the product
$\tilde{k}^2\langle\Psi(N)|\hat{H}_{\bf k}|\Psi(N)\rangle$, which appears
in the evaluation of the ground state energy (the factor $\tilde{k}^2$ coming from the Jacobian
in spherical coordinates in three dimensions), as a function of the dimensionless wavevector 
$\tilde{k}$, for the above two choices of the constants $c_{\bf k}$.
It can be seen from this plot that the solution (\ref{Eq:solck}) with nonzero $\tilde{\sigma}$ (solid line)
leads to a lower value of the ground state energy (\ref{Eq:expectation_Hk}) than the standard
solution of the single-mode theory with $\tilde{\sigma}=0$ from Eq. (\ref{Eq:resultck}), (dashed line).
Since the coefficients in Eqs. (\ref{Eq:resultck}) and (\ref{Eq:solck}) are 
quite different from one another, it follows that
expectation values of observables calculated with the coefficients $c_{\bf k}$
obtained by minimizing the expectation value of the full Hamiltonian $\hat{H}$ will be
quite different from the expectation values of the same observables calculated
using the usual Bogoliubov approximation where, for each value of ${\bf k}$, 
the expectation value of the single-mode Hamiltonian $\hat{H}_{\bf k}$ is minimized.

\begin{figure}[tb]
\includegraphics[width=8.09cm, height=5.5cm]{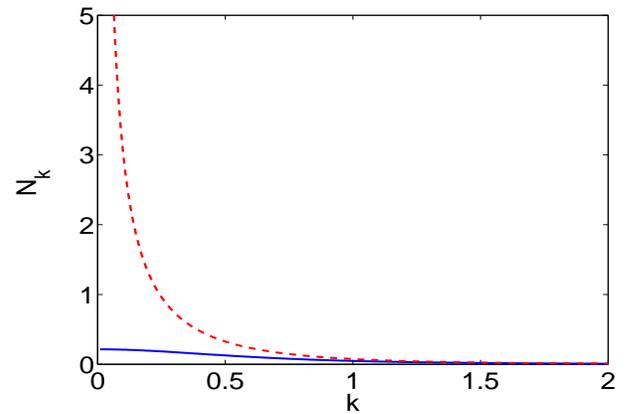}
\caption[]{Plot of the depletion of the condensate $N_{\bf k}$ as a function of the dimensionless wavevector
$\tilde{k}$ using the coefficients $c_{\bf k}$
(i) from the single-mode theory, Eq. (\ref{Eq:resultck}), (dashed line) 
and (ii) from Eq. (\ref{Eq:solck}) with $\tilde\sigma=0.39$ and $C_d=0.9762$ (solid line).
}\label{Fig:plotNk}
\end{figure}

As a first example, in Fig. \ref{Fig:plotNk}, we plot the depletion $N_{\bf k}$ of the condensate
$N_{\bf k}=\langle\Psi(N)|a_{\bf k}^\dagger a_{\bf k}|\Psi(N)\rangle
= c_{\bf k}^2/(1-c_{\bf k}^2)$ {\em vs.} wavevector $\tilde{k}$, 
with the solid line representing the result one obtains
using the coefficients $c_{\bf k}$ from Eq. (\ref{Eq:solck}), and the dashed line
representing the results one obtains using Eq. (\ref{Eq:resultck}).
As it can be seen, the two results are qualitatively very different, with $N_{\bf k}$ diverging
like $1/k^2$ as $k\to 0$ in the standard Bogoliubov theory (which is of course unphysical
for a system of fixed number of bosons $N$, and leads in one spatial dimension
to an infrared divergence of the total number of depleted bosons $N$), while,
on the contrary, when the improved coefficients $c_{\bf k}$ of Eq. (\ref{Eq:solck}) are
used, $N_{\bf k}$ is finite for all values of the wavevector $k$.

As a second example, we consider the energy to excite one boson from the condensate
to the single-particle state with wavevector ${\bf k}$. This is the quantity given by:
\begin{equation}
\Delta E_{\bf k} = \frac{\langle\Psi| a_0^\dagger a_{\bf k} \hat{H} a_0a_{\bf k}^\dagger|\Psi\rangle}
{\langle\Psi| a_0^\dagger a_{\bf k} a_0a_{\bf k}^\dagger|\Psi\rangle}
-\langle\Psi|  \hat{H} |\Psi\rangle
\end{equation}
In the standard formulation of Bogoliubov's theory, 
where $a_0$ and $a_0^\dagger$ are replaced by the c-number $\sqrt{N_0}$ 
(where $N_0$ is the number of bosons in the condensate), 
we obtain:
\begin{equation}
\Delta E_{\bf k} = n_Bv({\bf k})\Big[
\big(\tilde{k}^2 + 1\big)\frac{1+c_{\bf k}^2}{1-c_{\bf k}^2} 
-\frac{2c_{\bf k}}{1-c_{\bf k}^2}\Big].
\label{Eq:excBog}
\end{equation}
On the other hand, in the variational treatment of Eq. 
(\ref{Eq:fullPsi(N)}), where we keep an accurate
count of the number of bosons in the ${\bf k}=0$ state
as is done in the basis wavefunctions of Eq. (\ref{Eq:varBasis}), 
it can be shown \cite{Ettouhami} that the quantity $\Delta E_{\bf k}$
is given by:
\begin{equation}
\Delta E_{\bf k} = n_Bv({\bf k})\Big\{
\big[C_d^{-1}(\tilde{k}^2 + \tilde\sigma) + 1\big]\frac{1+c_{\bf k}^2}{1-c_{\bf k}^2} 
-\frac{2c_{\bf k}}{1-c_{\bf k}^2}\Big\}.
\label{Eq:excModBog}
\end{equation}
Using the expression of $c_{\bf k}$ given by Eq. (\ref{Eq:resultck}) in Eq. (\ref{Eq:excBog}) above,
one obtains the celebrated Bogoliubov spectrum $\Delta E_{\bf k}=\sqrt{\tilde{k}^2(\tilde{k}^2+2)}$,
which is gapless as $k\to 0$. Conversely, when the coefficients $c_{\bf k}$ of Eq. (\ref{Eq:solck}) are used
in Eq. (\ref{Eq:excModBog}), one obtains:
\begin{equation}
\Delta E_{\bf k} = n_Bv(k)\sqrt{Q^2(Q^2+2)}, \quad Q^2 \simeq C_d^{-1}(\tilde{k}^2 + \tilde{\sigma}),
\end{equation}
which has a finite gap as $k\to 0$, $\Delta E_{\bf k\to 0} = gn_BC_d^{-1}\sqrt{\tilde{\sigma}(\tilde{\sigma}+2C_d)}$. 
For the values of $n_Ba^3$, $\tilde\sigma$ and $C_d$ considered in this paper, we
obtain $\Delta E_{\bf k\to 0}= 0.98gn_B$, which is comparable to the value of the gap $gn_B$ predicted
by the standard Hartree-Fock method, and by Girardeau and Arnowitt in Ref. \onlinecite{Girardeau}.

{\em Conclusion.} To summarize, in this paper, we have argued that the decoupled way
in which the Hamiltonian $\hat{H}=\sum_{\bf k\neq 0} \hat{H}_{\bf k}$ is diagonalized in the standard formulation of
Bogoliubov's theory, where each and every momentum contribution $\hat{H}_{\bf k}$
is diagonalized separately, is not appropriate. Diagonalizing the total Hamiltonian
$\hat{H}$ directly leads to results that are {\em markedly} different
from the results of Bogoliubov's method.  More specifically, we find 
that the depletion of the condensate is smaller than what Bogoliubov's theory predicts,
and that the energy to excite a single boson from the condensate to the single-particle
state with wavevector ${\bf k}$ has a finite gap as $k\to 0$.
A more thorough analysis detailing further evidence in support of the above conclusions,
including a more detailed discussion of the elementary excitations of the 
full Hamiltonian $\hat{H}$, will be presented elsewhere \cite{Ettouhami}.


\begin{thebibliography}{99}


\bibitem{Bogoliubov1947} N.N. Bogoliubov, J. Phys. U.S.S.R. {\bf 5}, 71 (1947).

\bibitem{Lee1957} T.D. Lee and C.N. Yang, Phys. Rev. {\bf 105}, 1119 (1957).

\bibitem{LeeHuangYang1957} T.D. Lee, K. Huang and C.N. Yang, Phys. Rev. {\bf 106}, 1135 (1957).

\bibitem{Bruckner1957} K.A. Bruckner and K. Sawada, Phys. Rev. {\bf 106}, 1117 (1957).

\bibitem{Beliaev1958} S.T. Beliaev, Soviet Physics JETP {\bf 7}, 104 (1958); {\bf 7}, 289 (1958).

\bibitem{Hugenholtz1959} N.M. Hugenholtz and D. Pines, Phys. Rev. {\bf 116}, 489 (1959).

\bibitem{Sawada1959} K. Sawada, Phys. Rev. {\bf 116}, 1344 (1959).

\bibitem{Gavoret1964} J. Gavoret and P. Nozi\`eres, Ann. Phys. (New York) {\bf 28}, 349 (1964).

\bibitem{Hohenberg1965} P.C. Hohenberg and P.C. Martin, Ann. Phys. (New York) {\bf 34}, 291 (1965).

\bibitem{Popov1965} V.N. Popov, Soviet Physics JETP {\bf 20}, 1185 (1965).

\bibitem{Singh1967} K.K. Singh, Physica {\bf 34}, 285 (1967).

\bibitem{Cheung1971} T.H. Cheung and A. Griffin, Phys. Rev. A {\bf 4}, 237 (1971).

\bibitem{Szepfalusy1974} P. Sz\'epfalusy and I. Kondor, Ann. Phys. (New York) {\bf 82}, 1 (1974).

\bibitem{LeggettRMP} A.J. Leggett, Rev. Mod. Phys. {\bf 73}, 307 (2001).

\bibitem{Girardeau} M. Girardeau and R. Arnowitt, Phys. Rev. {\bf 113}, 755 (1959).

\bibitem{Ettouhami} A.M. Ettouhami (unpublished).



\end{thebibliography}
\end{document}